\documentclass[twocolumn,showpacs,floatfix,prl]{revtex4}

\usepackage{float}
\usepackage{graphicx}

\begin{document}

\title{Hexagonal Warping Effects in the Surface States of Topological Insulator Bi$_2$Te$_3$}

\author{Liang Fu}
\email{liangfu@physics.harvard.edu}
\affiliation{Department of Physics, Harvard University, Cambridge, MA 02138}

\begin{abstract}
A single two-dimensinoal Dirac fermion state has been recently observed on the surface of topological insulator Bi$_2$Te$_3$ 
by angle-resolved photoemission spectroscopy (ARPES). We study the surface band structure using $k \cdot p$ theory and find an unconventional
hexagonal warping term, which is the counterpart of cubic Dresselhaus spin-orbit coupling in rhombohedral structures. 
We show that this hexagonal warping term naturally explains the observed hexagonal snow-flake Fermi surface.  The strength of hexagonal warping is characterized by a single parameter, which is extracted from the size of the Fermi surface. 
We predict a number of testable signatures of hexagonal warping in spectroscopy experiments on Bi$_2$Te$_3$. 
We also explore the possibility of a spin-density wave due to strong nesting of the Fermi surface.
\end{abstract}

\maketitle

Recently a new state of matter called topological insulators has been observed 
in a number of materials\cite{hasanbisb, japanbisb, hasanbise, shenbite, hasanbite}. A topological insulator has a time-reversal-invariant band structure with nontrivial 
topological order, which gives rise to gapless surface states bound to the sample boundary\cite{fkm, moorebalents, roy}.  
The two-dimensional surface band has a unique Fermi surface that encloses an {\it odd} number of Dirac points in the surface Brillouin zone\cite{fkm}, 
which is prohibited in  conventional materials by fermion doubling theorem\cite{doubling}. 
Soon after the theoretical prediction\cite{fukane}, the semiconducting alloy Bi$_x$Sb$_{1-x}$ was found to be a topological insulator having
a Dirac surface band as well as other electron and hole pockets\cite{hasanbisb}. Subsequently, a family of materials Bi$_2$X$_3$ (X=Se and Te) was found to 
be topological insulators with a  {\it single} Dirac-fermion surface state\cite{hasanbise, shenbite, hasanbite}. 
The observation of an undoubled Dirac fermion is not only of great conceptual interest but also paves the way for 
studying unusual electromagnetic properties\cite{fukane, qi} and 
realizing topological quantum computation\cite{fkmajorana}. Therefore surface states of Bi$_2$X$_3$ are being intensively studied in transport and spectroscopy experiments\cite{ong, mbe}. 

In this work, we study the electronic properties of surface states in Bi$_2$Te$_3$ using $k\cdot p$ theory. 
Our motivation is to understand the shape of Fermi surface observed in recent ARPES experiments\cite{shenbite, hasanbite}, reproduced in Fig.1.
By considering the crystal symmetry of Bi$_2$Te$_3$, we find an unconventional hexagonal warping term in the surface band structure, 
which is the counterpart of cubic Dresselhaus spin-orbit coupling in rhombohedral structures. 
This hexagonal warping term naturally explains the snow-flake shape of the Fermi surface, and its magnitude is extracted from the size of Fermi surface. 
We predict that hexagonal warping of the Fermi surface should have important effects in several spectroscopy experiments. 
Finally, we observe that Fermi surface of Bi$_2$Te$_3$ is nearly a hexagon with strong nesting for an appropriate range of surface charge density. 
This motivates us to explore theoretically a possible spin-density wave (SDW) phase. We discuss various types of SDW order in a Landau-Ginzburg theory.






Bi$_2$Te$_3$ has a rhombohedral crystal structure with space group $R3\bar{m}$. In the presence of a [111] surface, the symmetry of the crystal  
is reduced to $C_{3v}$, which consists of a three-fold rotation $C_3$ around the trigonal $z$ axis and a mirror operation $M: x \rightarrow -x$ where $x$ is in $ \Gamma  K$ direction. 
Two surface bands are observed to touch at  the origin of the surface Brillouin zone $\Gamma$. The degeneracy is protected by time-reversal symmetry and the doublet $| \psi_{\uparrow, \downarrow} \rangle$ form a Kramers pair. 
We choose a natural basis for the doublet according to total angular momentum $J=L+S=\pm 1/2$ so that 
$C_3$ is represented as $e^{-i \sigma_z \pi/3 }$. Since $M^2=-1$ for spin $1/2$ electron and $M C_3 M^{-1}=C_3^{-1}$,  the mirror operation can be represented as $M= i\sigma_x$ by
defining the phase of $| \psi_{\uparrow, \downarrow} \rangle$ appropriately. The anti-unitary time reversal operation $\Theta$ is represented by $i\sigma_y K$ ($K$ is complex conjugation) and commutes with both $M$ and $C_3$. Here  the pseudo-spin $\sigma_i$ is proportional to electron's spin: $\langle s_z \rangle \propto \langle \sigma_z \rangle$ and $\langle s_{x,y} \rangle \propto  \langle \sigma_{x,y} \rangle$.
 
\begin{figure}
\centering
\includegraphics[width=3in]{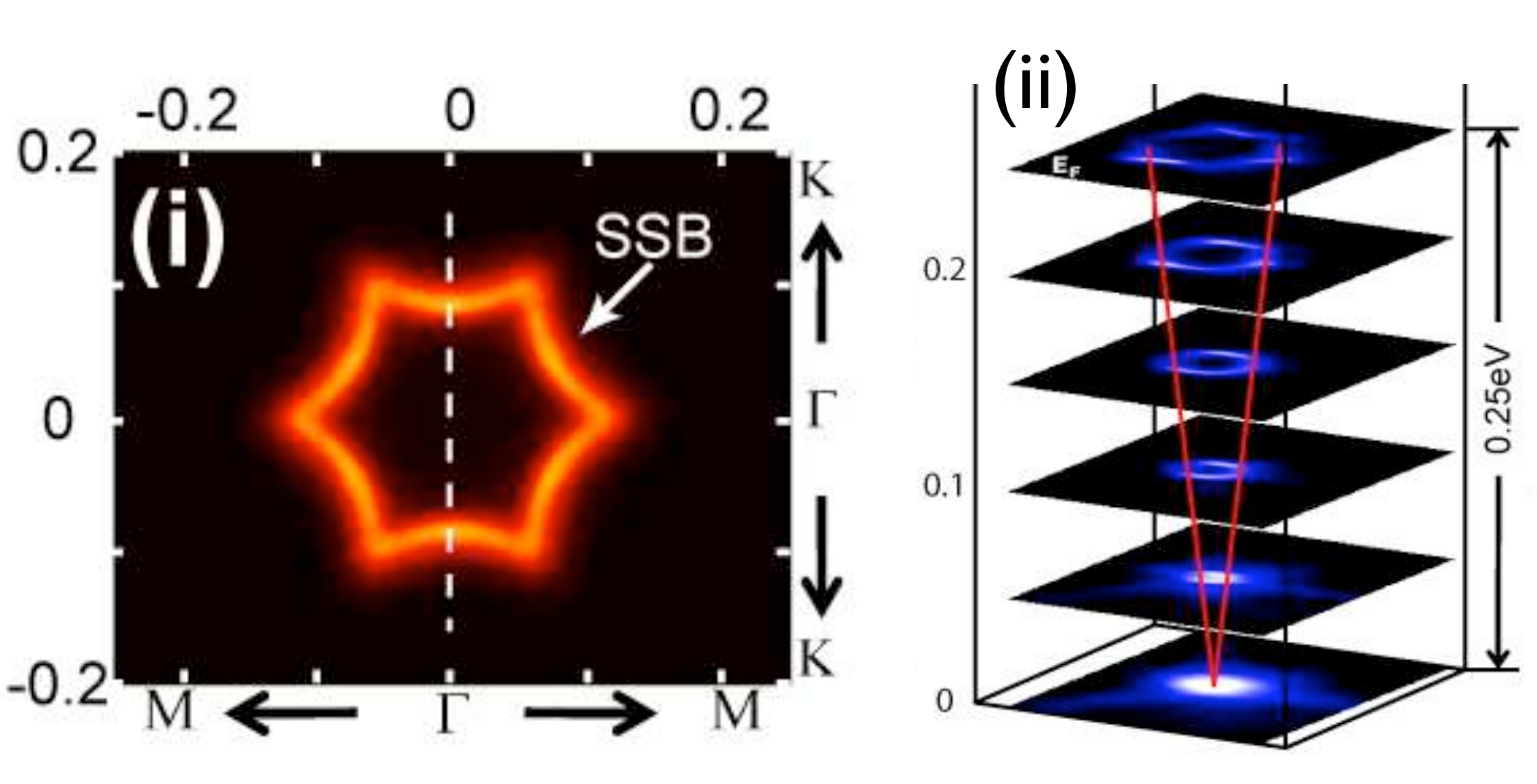}
\caption{(i) snow-flake like Fermi surface of the surface states on $0.67\%$ Sn-doped Bi$_2$Te$_3$ observed in ARPES. (ii) a set of constant energy contours at different energies. From \cite{shenbite}. Reprinted with permission from AAAS.}
\end{figure}

The Kramers doublet is split  away from $ \Gamma$ by spin-orbit interaction. We study the surface band structure near $\Gamma$ 
using $k\cdot p$ theory. To lowest order in $k$, the $2 \times 2$ effective Hamiltonian reads
$H_0= v (k_x \sigma_y - k_y \sigma_x)$,
which describes an isotropic 2D Dirac fermion. The form of $H_0$ is strictly fixed by symmetry. In particular, the Fermi velocity $v$ in $x$ and $y$ directions are equal because of the $C_3$ symmetry. 
The Fermi surface of $H_0$ at any Fermi energy is a circle. However, the Fermi surface observed in ARPES, reproduced in Fig.1(i), is non-circular but snow-flake like: it has relatively sharp tips extending along six $\Gamma M$ directions and curves inward in between. Moreover, as shown in Fig.1(ii) (we refer the reader to the original work\cite{shenbite} for better resolution), the shape of constant energy contour is energy-dependent, evolving from a snow-flake at $E=0.25eV$ to a hexagon and then to a circle near the Dirac point. Throughout this paper, energy is measured with respect to the Dirac point.

The observed anisotropic Fermi surface can only be explained by higher order terms 
in the $k \cdot p$ Hamiltonian $H(\vec{k})$ that break the emerging $U(1)$ rotational symmetry of $H_0$.
The form of $H(\vec{k})$ is highly constrained by crystal and time reversal symmetry. Under the operation of $C_3$ and $M$, momentum and  spin transform as follows:
\begin{eqnarray}
C_3 &:& k_{\pm} \rightarrow  e^{\pm i2\pi/3} k_{\pm},  \;\;\;  \sigma_{\pm} \rightarrow  e^{\pm i2\pi/3} \sigma_{\pm}, \;\;\; \sigma_z \rightarrow \sigma_z \nonumber \\
M &:& k_+ \leftrightarrow -k_-,  \;\;\; \sigma_x \rightarrow \sigma_x, \;\;\; \sigma_{y,z} \rightarrow - \sigma_{y,z}, 
\label{c3m}
\end{eqnarray}
where $k_{\pm}= k_x \pm i k_y$ and $\sigma_{\pm} = \sigma_x \pm i \sigma_y$. $H(\vec{k})$ must be invariant under (\ref{c3m}). 
In addition, time reversal symmetry gives the constraint
\begin{equation}
H(\vec{k})=\Theta H(-\vec{k}) \Theta^{-1}= \sigma^y H^*(-\vec{k}) \sigma^y. \label{TR}
\end{equation}
We then find that $H(\vec{k})$ must take the following form up to third order in $\vec{k}$:
\begin{equation}
H(\vec{k})= E_0(k) + v_k (k_x \sigma_y - k_y \sigma_x)  +  \frac{ \lambda}{2} (k_+^3 + k_-^3) \sigma_z, \label{kpH} 
\end{equation}
where $E_0(k) = k^2/ (2m^*)$ generates particle-hole asymmetry and the Dirac velocity $v_k=v(1+\alpha k^2)$ contains a second-order correction. 
The last term in (\ref{kpH}), which we call $H_w$, is most important. Unlike the other terms, $H_w$ is only invariant under three-fold rotation (as the Bi$_2$Te$_3$ crystal structure does) and therefore is solely responsible for the hexagonal distortion of the otherwise circular Fermi surface. We note that $H_w(\vec{k})$ vanishes in mirror-symmetric direction $\Gamma M$, because $\sigma^z$ is odd under mirror, and $H_w(\vec{k})$ reaches maximum along $\Gamma K$.  
The surface band dispersion of $H(\vec{k})$ is
\begin{equation}
E_{\pm}(\vec{k}) = E_0(k) \pm \sqrt{v_k^2 k^2 + \lambda^2 k^6 \cos^2(3\theta)}. \label{dispersion}
\end{equation} 
Here $E_{\pm}$ denote the energy of upper and lower band, and $\theta$ is the  azimuth angle of momentum $\vec{k}$ with respect to the x axis ($\Gamma K$). Although the Hamiltonian $H$ is three-fold invariant, the band structure is six-fold symmetric under $\theta \rightarrow \theta + 2\pi/6$ because of time reversal symmetry.

The hexagonal warping term $H_w$ describes cubic spin-orbit coupling at the surface of rhombohedral crystal systems, and to the best of our knowledge, it has not  been reported before. It is instructive to compare $H_w$ with the well-studied trigonal warping in graphene\cite{graphenewarping}. Although graphene's band structure also has Dirac points and its $k \cdot p$ Hamiltonian $H_K(\vec{k})$ has $C_{3v}$ symmetry, the warping term in graphene is of a completely different form. This is because time reversal operation $\Theta$ acts differently for spin $1/2$ ($\Theta=i\sigma_y K$) and spinless fermions ($\Theta=K$).  In graphene time reversal symmetry takes the latter form, and together with inversion symmetry, leads to $H_K(\vec{k})=\tau_x H^*_K(\vec{k})\tau_x$ ($\tau_z=\pm 1$ denote two sublattice), as opposed to its partner Eq.(\ref{TR}) in Bi$_2$Te$_3$. As a result, a different trigonal warping term $ (k_+^2 \tau_+ + k_-^2\tau_-)$ is symmetry-allowed in graphene. 


   


\begin{figure}[htp]
\centering
\includegraphics[width=3in]{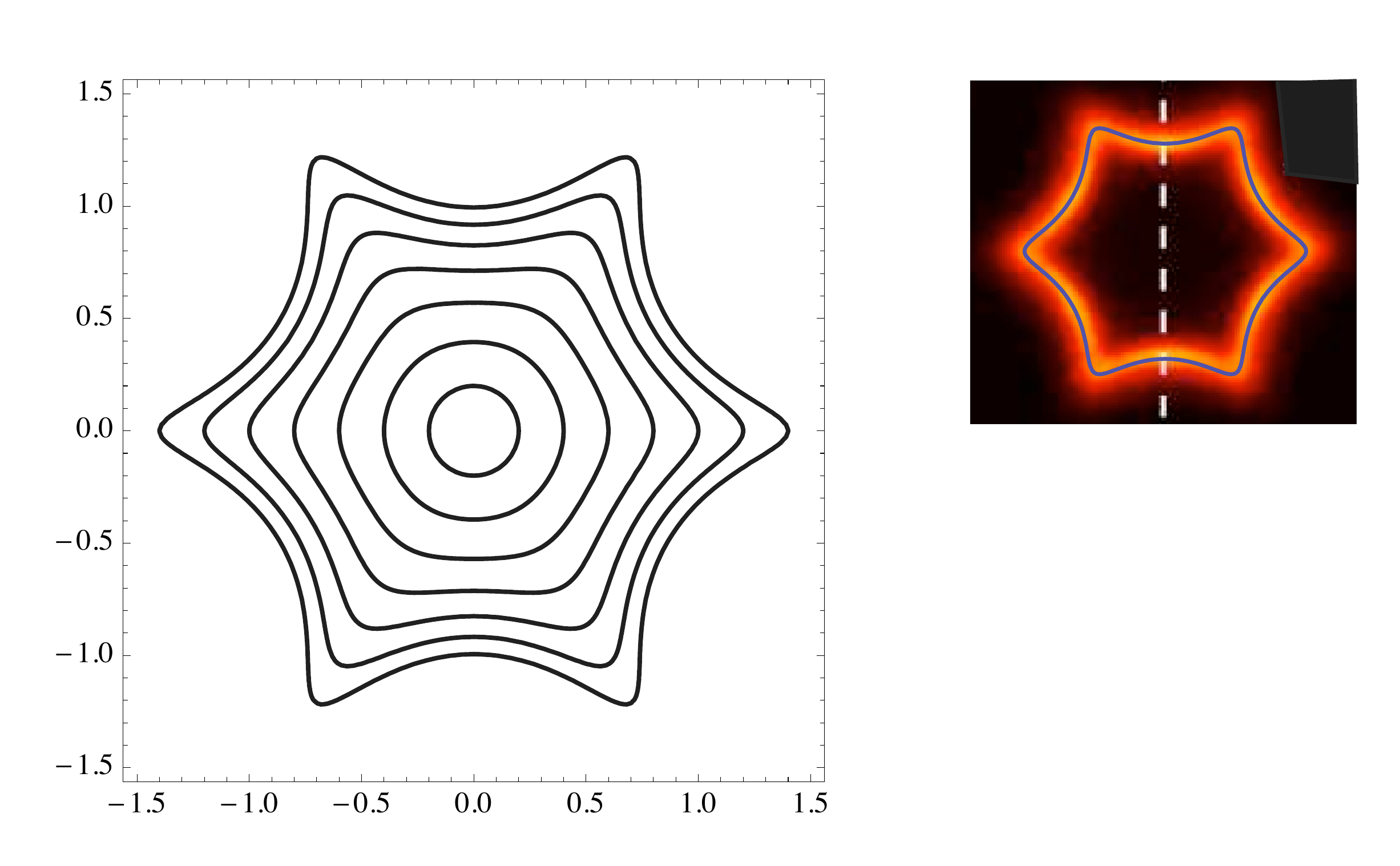}
\caption[FermiCircle]{Left: constant energy contour of $H(\vec{k})$. $k_x$ and $k_y$ axis are in the unit of $\sqrt{v/\lambda}$. Right: constant energy contour at $E=1.2E^*$ is superimposed on the Fermi surface of Bi$_2$Te$_3$. 
}
\end{figure}

We now show that $H_w$ naturally explains the observed energy-depdent shape of Fermi surface in Bi$_2$Te$_3$.
Using (\ref{dispersion}) we plot a set of constant energy contours of $H(\vec{k})$ for $0<E<2E^*$  in Fig.2, where $E^* \equiv v/a$ and $a \equiv \sqrt{\lambda/v}$ are the characteristic energy and length scale introduced by hexagonal warping. 
For simplicity, we have discarded $E_0$ and the quadratic correction to velocity, since they do not change the shape of Fermi surface significantly.
By plotting $k_x$ and $k_y$ axis in the unit of $\sqrt{v/\lambda}$, Fig.2 is obtained with {\it no free parameter}. 
As shown in the figure, Fermi surface starts to deviate considerably from a circle and becomes more hexagon-like around $E=0.55E^*$. 
When $E > E_c \equiv \sqrt{7}/6^{3/4} E^* \approx 0.69 E^*$, the edge of the hexagon curves inward so that Fermi surface ceases to be purely convex. 
As $E$ further increases, rounded tips starts to develop at the vertices of the hexagon, which eventually become sharper making the Fermi surface snow-flake like. 
The evolution of Fermi surface with respect to energy matches well with the ARPES result shown in Fig.1. Moreover, it follows from (\ref{dispersion}) that the vertices of the hexagon---where the Fermi surface extends outmost---{\it always} lie along $\Gamma M$ independent of the sign of $\lambda$, in agreement with ARPES data.

Comparing the set of Fermi surfaces in Fig.2a with the real Fermi surface in $0.67\%$ Sn-doped Bi$_2$Te$_3$ (Fig.1), we find the Fermi surface at $E_F=1.2E^*$  is almost {\it identical} to the one measured in ARPES, as shown by superimposing the two in Fig.2b. By fitting the theoretical value of Fermi momentum along $\Gamma M$ ($1.2/a$) to the experimental one ($0.11\AA^{-1}$), we find $a = 10.9 \AA$. Using the measured Fermi velocity $v=2.55eV \cdot \AA$, we obtain the magnitude of the hexagonal warping term: $\lambda = 250eV \cdot \AA^3$. From that we find $E^*=0.23eV$, and $E_F=1.2E^*=0.28eV$ which agrees fairly well with the measured Fermi energy $0.25eV$ (shown in Fig.1.ii). %
The quantitative agreement between theory and experiment suggests that the Hamiltonian (\ref{kpH}) describes the surface band structure of Bi$_2$Te$_3$ quite well in a wide energy window at least up to $0.25eV$. 
As an independent check of the theory, we consider the non-linear correction to surface band dispersion near $\Gamma$. (\ref{dispersion}) predicts that the leading order correction due to $H_w$ starts at {\it fifth} order in $k$ and is angle-dependent:
\begin{equation}
\delta E(k, \theta) = v a^4 k^5 \cos^2(3 \theta)/2. \label{correction}
\end{equation}
Since the surface band dispersions along $\Gamma K$ and $\Gamma M$ directions have been measured in ARPES\cite{shenbite, hasanbite}, (\ref{correction}) can be tested by fitting to $E_{\Gamma M}(k)-E_{\Gamma K}(k)$, which also gives an independent way of obtaining $\lambda$. 
The two other parameters $m^*$ and $\alpha$ in $H(\vec{k})$ can also be extracted by a careful fitting to the band dispersion.     
 

From now on, we predict a variety of important effects of hexagonal warping in Bi$_2$Te$_3$. 
First,  because $H_w$ couples to $\sigma_z$,  the spin polarization of surface states should have an out-of-plane component $s_z   \propto \langle \sigma_z \rangle$. 
Since spin polarization along $\Gamma M$ has been found to be almost $100\%$ polarized in a very recent ARPES experiment\cite{hasanbite}, 
we conclude that the the doublet $| \psi_{\uparrow,\downarrow} \rangle$ at $\Gamma$ are almost pure spin eigenstates, i.e., $s_z \approx \langle \sigma_z \rangle$, which agrees with a theoretical band structure calculation\cite{bitetheory}. 
$s_z$ is then calculated from (\ref{kpH}): $s_z  = \cos(3\theta)/ \sqrt{\cos^2(3\theta)+1/(ka)^4}$. The out-of-plane spin polarization is momentum-dependent and can reach as high as $60\%$ of the full polarization along $\Gamma K$ for the Fermi surface in Fig.1. 
We hope this pattern of out-of-plane spin polarization can be tested in future spin-resolved ARPES.  


Second, hexagonal warping gives a novel mechanism for opening up an energy gap at the Dirac point. 
Consider an in-plane magnetic field $B_{\parallel}$, which only couples to the spin $H^{\parallel}_{Zeeman} = 
g_{\parallel} \vec{B}_{\parallel} \cdot \vec{\sigma}$. 
From (\ref{kpH}) we find the Dirac point is shifted away from $\Gamma$ to $\vec{k}^* \equiv g_{\parallel} \hat{z} \times  \vec{B}_{\parallel}/v$. In addition to that, a mass term is generated at $\vec{k}^*$: $M\sigma_z= (g_{\parallel} B_{\parallel})^3  \sin(3\varphi) \sigma_z / {E^*}^2$ ($\varphi$ is the angle between $\vec{B}_{\parallel}$ and $\Gamma K$), which opens up an energy gap. When the Fermi energy is tuned, e.g. by doping\cite{shenbite, hasanbite}, to lie within the gap, the insulating state at the surface realizes quantum Hall effect without Landau levels\cite{haldane}. 


\begin{figure}[htp]
\centering
\includegraphics[width=3in]{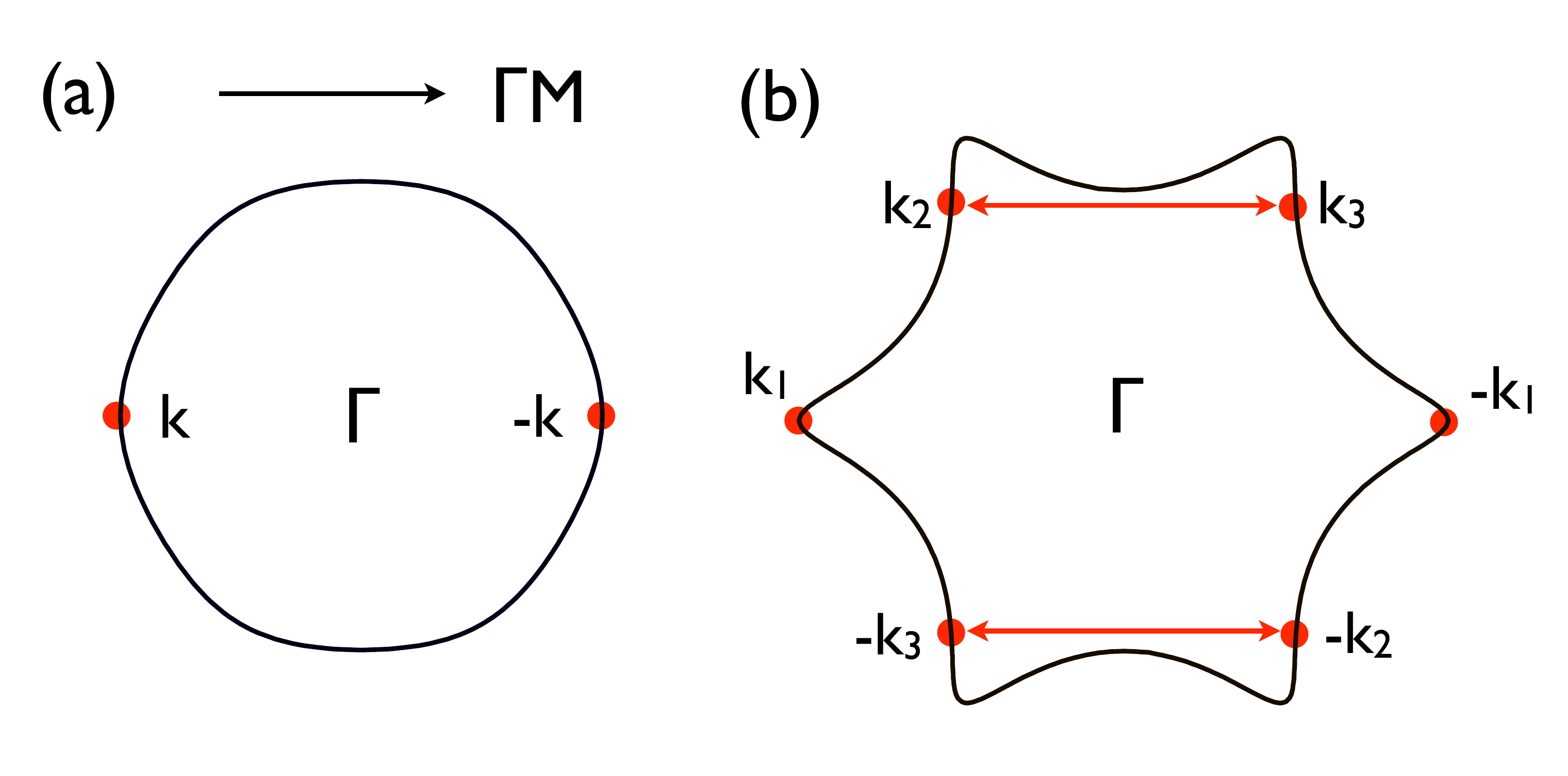}
\caption{Illustration of scattering processes due to a point defect that causes the oscillation of LDOS.
In a given direction $\hat{x}$ along $\Gamma M$, the oscillation is dominated by scattering between stationary points marked by dot, where the Fermi velocity is parallel to $\hat{x}$.   
(a) a convex constant energy contour has a single pair of stationary point at $\vec{k}$ and $-\vec{k}$. (b) a non-convex constant energy contour has three pairs of stationary points. Intra-pair scatterings in (a) and (b) are forbidden by time reversal symmetry. But inter-pair scatterings in (b), for example, those between $k_2$ and $k_3$, are allowed. Therefore, the LDOS oscillation at leading order is absent in (a) but exists in (b). 
}
\end{figure}

Third, hexagonal warping of Fermi surface has drastic effects on the Friedel oscillation of local density of states (LDOS) around a nonmagnetic point defect in STM. The LDOS oscillation {\it at a fixed energy} decays algebraically as a function of distance away from the defect. In a normal 2D metal, the leading order ($1/x$) decay of LDOS in a given direction $\hat{x}$ comes from scattering between states at ``stationary points'' on the Fermi surface, where the Fermi velocity is parallel to $\hat x$\cite{roth}. 
For a convex constant energy contour below $E_c$ as shown in Fig.3a, only a single pair of stationary points exist at $\vec k$ and $-\vec k$. However, since the two states at $\vec k$ and $-\vec k$ here carry opposite spins, scattering between them is forbidden by time reversal symmetry---a fundamental property of surface states on a topological insulator. The LDOS oscillation then vanishes at leading order. Now consider a non-convex constant energy contour above $E_c$. As shown in Fig.3b, {\it multiple} pairs of stationary points exist. Since inter-pair scattering is still allowed, the LDOS oscillation will be restored at leading order. Therefore according to the convexity of constant energy contour, two types of Friedel oscillation patterns should appear at different ranges of bias voltage. 
   
In the last part of this work, we explore the possibility of a spin-density wave (SDW) phase on the surface of Bi$_2$Te$_3$. 
We note that the Fermi surface is nearly a hexagon for $0.55E^*<E<0.9E^*$. The almost flat pieces on the edges of the hexagon leads to strong nesting at wave-vectors $Q_i= 2k_F {\bf e}_i, i=1,...3$, where $k_F {\bf e}_i$ is the Fermi momentum in three equivalent $\Gamma K$ directions. A density-wave ordered phase may then exist at a finite interaction strength. Since the surface states at $\vec{k}$ and $-\vec{k}$ have opposite spins, a charge-density wave cannot connect them and is thus disfavored. We are therefore motivated to consider possible SDW phases. 

We now discuss the phase diagram of SDW in a Landau-Ginzburg theory based on general symmetry considerations. We define the order parameters of the SDW as follows,
\begin{eqnarray}
\phi_{i \parallel} &=& \sum_k  \langle c_{k+Q_i }^\dagger   {\bf e}_i \cdot \vec{\sigma} c_{k} \rangle, \nonumber \\
\phi_{i \perp} &=& i \sum_k \langle  c_{k+Q_i }^\dagger (\hat{z} \times {\bf e}_i ) \cdot \vec{\sigma} c_{k} \rangle \nonumber \\
\phi_{iz} &=& i \sum_k \langle c_{k+Q_i }^\dagger \sigma^z c_{k} \rangle  
\end{eqnarray}
where $c_{k}^\dagger=(c_{\uparrow k}^\dagger, c_{\downarrow k}^\dagger)$ are electron creation operators. 
For each $Q_i$, we have chosen a local frame for in-plane spin components labeled by $\parallel$ and $\perp$ which are parallel and perpendicular to $Q_i$ respectively. The order parameters thus defined transform nicely under the operations of rotation, mirror, time reversal and translation:
\begin{eqnarray}
C_3: \; & &\phi_{i \mu} \rightarrow   \phi_{i +1,\mu} \nonumber \\
\Theta: \;  & & \phi_{i, \mu} \rightarrow - \phi_{i,\mu}   \nonumber \\
M_x: \; & & \phi_{1\mu} \leftrightarrow \phi^*_{1\mu}, \;  \phi_{2\mu} \leftrightarrow \phi^*_{3\mu},  \nonumber \\
T_d: \; & & \phi_{i \mu} \rightarrow e^{i Q_i \cdot d} \phi_{i\mu}, \;\; \mu=\parallel, \perp, z 
\label{symmetry}
\end{eqnarray}
We remark that because spin and momentum are locked by spin-orbit coupling, there is no $SU(2)$ symmetry for spin alone. 
Thanks to the appropriate choice of order parameters, symmetry operations (\ref{symmetry}) only act in the space of ordering wave vectors labelled by $i$ index.  

The Landau free energy $F$ must be invariant under these symmetry operations. Only terms with even powers of $\phi_{i\mu}$ can exist because of time reversal symmetry. At second order, we have
\begin{equation}
F_2= \frac{1}{2}  \chi_{\mu \nu} \sum_{i=1}^3 \phi_{i\mu}^*\phi_{i\nu}, \label{f2}
\end{equation}
where the susceptibility matrix $\chi_{\mu \nu}$ is real and symmetric because of mirror symmetry. $\chi_{\mu \nu}$ is positive definite in the normal state. When  the temperature is lowered below $T_c$, one of the eigenvalues of $\chi_{\mu \nu}$ first becomes negative, and the surface undergoes a transition to a SDW. The  spin configuration is then determined by  the corresponding eigenvector $v_{\mu}$. 
For example, for a stripe SDW along $x$ direction, $\vec{S}(x,y) =(v_{\parallel} \cos(Qx), v_{\perp} \sin(Qx), v_z \sin(Qx))$ with an appropriate choice of origin.

The free energy (\ref{f2}) to second order has an emerging $U(3)$ symmetry $\phi_{i\mu} \rightarrow U_{ij} \phi_{j\mu}$. So single- and multiple-$Q$ SDWs are degenerate. We now show that higher order terms in $F$ break the $U(3)$ symmetry and picks out a particular spatial ordering pattern. For that purpose, it is convenient to write  $\phi_{i\mu}=\xi_i v_{\mu}, \sum_i |\xi_i|^2=1$ and use $\xi_i$ as a new set of order parameters, which also transforms according to (\ref{symmetry}). At fourth order, we find an anisotropy term $F_4 = u \sum_{i=1}^3 | \xi_i |^4$. The sign of $u$ determines the relative weight of $\phi_i$ in the ordered phase. For $u<0$, only one of $\phi_i$, say $\xi_1$, is nonzero. The resulting SDW forms a one-dimensional stripe, which breaks $C_3$ but is invariant under mirror symmetry. 
For $u>0$, $|\xi_1|=|\xi_2|=|\xi_3|$ in the ordered phase, so that SDW forms a two-dimensional lattice. Each individual phase of $\xi_i$ depends on the choice of origin. Only the global phase of $\xi_1\xi_2\xi_3$ is gauge invariant and is fixed by the sixth-order term of the form $ C (\phi_1 \phi_2 \phi_3)^2 + C^*  (\phi_1^* \phi_2^* \phi_3^*)^2$ in $F$.  



{\it Note added}: During the final stage of this work, we learned that Alpichshev et al.\cite{stmbite} imaged with STM the standing wave of surface states on Bi$_2$Te$_3$ near a line defect (instead of a point defect considered in this work). The LDOS oscillation was found to exist in the energy range with snow-flake like constant energy contour, but strongly suppressed in the range with circular constant energy contour. This supports our explanation of the correlation between LDOS oscillation and convexity of constant energy contour. 

This work was initiated at University of Pennsylvania. We thank Charlie Kane for inspiring discussions. We are indebted to Bertrand Halperin for many insightful discussions and helpful comments on the manuscript. We thank Zahid Hasan and David Hsieh for numerous discussions on ARPES, as well as Aharon Kapitulnik, Anton Akhmerov and especially Cenke Xu for useful conversations. This work was supported by the Harvard Society of Fellows and NSF grant DMR-0605066.

 
%

\end{document}